\documentclass{ifacconf}

\usepackage{graphicx}      
\usepackage{natbib}        
\usepackage{bm}
\usepackage{amsmath}
\allowdisplaybreaks
\usepackage{amssymb}
\usepackage{xcolor}
\usepackage{mathtools}
\usepackage{booktabs}
\usepackage{multirow}
\usepackage{algorithm}
\usepackage{algpseudocode}

\newcommand{\bx}{\bm{x}}
\newcommand{\by}{\bm{y}}
\newcommand{\blam}{\bm{\lambda}}

\begin{document}
\begin{frontmatter}

\title{A Hybrid Reinforcement and Self-Supervised Learning Aided Benders Decomposition Algorithm\thanksref{ifacnotice}\thanksref{footnoteinfo}} 

\thanks[ifacnotice]{© 2026 Agyeman, Li, Mitrai, Daoutidis. This work has been accepted to IFAC World Congress 2026 for publication under a Creative Commons Licence CC-BY-NC-ND.}
\thanks[footnoteinfo]{The partial financial support of NSF CBET (award number 2313289) is gratefully acknowledged. IM would like to acknowledge financial support from the McKetta Department of Chemical Engineering.}

\author[UMN]{Bernard T. Agyeman} 
\author[UMN]{Zhe Li} 
\author[UTAustin]{Ilias Mitrai}
\author[UMN]{Prodromos Daoutidis}

\address[UMN]{Department of Chemical Engineering and Materials Science, University of Minnesota, Minneapolis, MN 55455, United States.}

\address[UTAustin]{McKetta Department of Chemical Engineering, The University of Texas at Austin, Austin, TX 78712, United States.}

\begin{abstract} 
We propose a hybrid reinforcement and self-supervised learning framework for accelerating generalized Benders decomposition (GBD). In this framework, a graph-based reinforcement learning agent operates on a bipartite representation of the master problem and, together with a verification mechanism, determines the integer variable assignments that solve the master problem. These assignments are then used as inputs to a KKT-informed neural network, trained via self-supervision to predict primal–dual solutions that approximately satisfy the Karush–Kuhn–Tucker conditions of the subproblem. The predicted solutions are used to construct Benders cuts directly. The framework is evaluated on a mixed-integer nonlinear programming case study, where it achieves a 57.5\% reduction in solution time relative to classical GBD while consistently recovering optimal solutions across all test instances.
\end{abstract}


\end{frontmatter}

\section{Introduction}
Generalized Benders decomposition (GBD)~\citep{geoffrion1972generalized} is a widely used algorithm for solving mixed-integer nonlinear programs (MINLPs). It decomposes the original problem into a master problem that handles the integer variables and a subproblem that handles the continuous variables. The two problems are solved sequentially and coordinated via Benders cuts, yielding a sequence of lower and upper bounds that, under convexity assumptions on the subproblem, converge to the optimal solution. GBD has been successfully applied in mixed-integer optimal control~\citep{menta2020learning}, model predictive control~\citep{warrington2019learning}, and locomotion planning~\citep{ren2025accelerating}.

Despite its effectiveness, a straightforward implementation of GBD can be computationally demanding. Its performance depends on the complexity of both the master problem and the subproblem, as well as the quality of the generated Benders cuts. To address these challenges, several acceleration strategies have been proposed, including exploiting problem structure~\citep{saharidis2011initialization}, approximating the solutions of the master problem and the subproblem~\citep{mitrai2024computationally}, and cut management techniques~\citep{su2015computational}. In this work, we focus on approaches that leverage approximate solutions of both the master problem and the subproblem.

The master problem is a mixed-integer program whose complexity increases as cuts are added during the solution process. Classical strategies to reduce its computational cost include solving it approximately using heuristics, particularly in early iterations where the relaxation is weak~\citep{geoffrion1974multicommodity, raidl2014speeding, sohn2012hybrid}. For the subproblem, the presence of nonlinear constraints can also lead to significant computational challenges. This has motivated approximate solution strategies, such as solving the subproblem to local optimality while still generating valid cuts~\citep{zakeri2000inexact}.

Machine learning (ML) has emerged as a promising tool for accelerating optimization algorithms by learning surrogates that approximate expensive computations or partially replace classical solvers~\citep{bengio2021machine, mitrai2025accelerating}. ML-based acceleration methods are particularly effective when solving families of related optimization problems under varying parameter realizations, where information from previously solved instances can be leveraged to guide the efficient solution of new instances. In the context of GBD, prior work has applied classification to identify valuable cuts~\citep{jia2021benders}, active learning to warm-start the master problem~\citep{mitrai2024takingb}, supervised learning to approximate subproblem solutions and cuts~\citep{mitrai2024computationally}, and reinforcement learning (RL) to control inexactness or guide solver decisions~\citep{li2025learning}.

In our recent work, we developed a graph-based RL agent that operates on a bipartite representation of the master problem and provides candidate assignments for the integer variables. A verification mechanism was introduced to ensure that the absence of feasibility and optimality guarantees in the agent’s predictions does not compromise the finite convergence of the GBD algorithm~\citep{agyeman2026graph}.While these approaches demonstrate the potential of ML for accelerating GBD, they typically focus on a single component of the algorithm. In this paper, we investigate a unified framework that leverages learning-based surrogates for both the master problem and the subproblem to achieve improved computational performance.

To this end, we build on the graph-based agent for the master problem and introduce a Karush–Kuhn–Tucker (KKT)-informed neural network (KINN) as a surrogate for the subproblem. For a fixed value of the integer variables, the subproblem solution can be characterized via the KKT conditions. The KINN is trained via a self-supervised loss that enforces these conditions, enabling it to predict primal–dual solutions that approximately satisfy them and support the direct construction of Benders cuts. Compared to existing neural network-based approaches for approximating KKT systems in optimization problems (e.g.,~\citep{nellikkath2022physics, chen2024physics}), the proposed formulation incorporates a tailored loss function and network architecture suited to the structure of the subproblem in GBD.

In the proposed framework, the graph-based agent and the KINN are coordinated within the GBD iterations. The agent generates assignments for the integer variables, which are processed through a verification mechanism to ensure feasibility with respect to the master problem and to preserve the correctness of the resulting bound updates at each iteration. The verified assignments are then used as inputs to the KINN to obtain approximate primal–dual solutions, which are used to construct Benders cuts. The effectiveness of the approach is demonstrated on a benchmark MINLP case study and compared against classical GBD and single-surrogate variants.

\section{Preliminaries}
\subsection{Problem formulation}
We consider a subclass of MINLPs in which the integer variables enter linearly in both the cost function and the constraints, while the continuous variables appear nonlinearly:
\begin{align}\label{eq:minlp_form}
	\min_{\bm{x},\, \bm{y}} \; &  f(\bm{x}) + \bm{e}^\top \bm{y} \\
	\text{s.t.} \;& \bm{h}(\bm{x}) + \bm{A}\bm{y} = \bm{0}, \\
	& \bm{g}(\bm{x}) + \bm{B}\bm{y} \leq \bm{0}, \\
	& \bm{K}\bm{y} - \bm{b} \leq 0, \label{eq:pure_binary_constraints} \\
	& \bm{x} \in \mathcal{X}, \quad \bm{y} \in \{0,1\}^m, \nonumber
\end{align}
where \( \bm{x} \in \mathbb{R}^n \) are continuous variables, \( \bm{y} \in \{0,1\}^m \) are binary variables, \( \bm{K} \in \mathbb{R}^{o \times m} \), \( \bm{b} \in \mathbb{R}^o \), \( \bm{e} \in \mathbb{R}^m \), \( \bm{A} \in \mathbb{R}^{p \times m} \), and \( \bm{B} \in \mathbb{R}^{q \times m} \). The set $\mathcal{X}$ is defined as
\[
\mathcal{X} \coloneqq \left\{ \bm{x} \in \mathbb{R}^n \,\middle|\, \bm{E}\bm{x} \leq \bm{d} \right\},
\]
where \( \bm{E} \in \mathbb{R}^{l \times n} \), \(\bm{d} \in \mathbb{R}^l\). The functions \( f \), \( \bm{h} \), and \( \bm{g} \) are continuous, differentiable, and convex. Inequality~\eqref{eq:pure_binary_constraints} represents constraints involving only the binary variables and are referred to as pure binary constraints.

\subsection{Generalized Benders decomposition}
GBD~\citep{geoffrion1972generalized,floudas1995nonlinear} proceeds by iteratively refining upper and lower bounds on the optimal solution of the original MINLP. At iteration $k$, the subproblem $\mathcal{S}(\bm{y}^k)$ is defined as
\begin{equation}
	\begin{aligned}\label{eq:minlp_sp}
		Z(\bm{y}^k) = \min_{\bm{x}} \; & f(\bm{x}) + \bm{e}^\top \bm{y}^k  \\
		\text{s.t.} \;& \bm{h}(\bm{x}) + \bm{A}\bm{y}^k = \bm{0}, \\
		& \bm{g}(\bm{x}) + \bm{B}\bm{y}^k \leq \bm{0}, \\
		& \bm{x} \in \mathcal{X}.
	\end{aligned}
\end{equation}

For simplicity, we assume that the subproblem is feasible for all $\bm{y}$. The subproblem objective value provides an upper bound (UBD) on the optimal value of the original MINLP. The original MINLP can then be equivalently expressed as
\begin{equation}
	\begin{aligned}\label{eq:minlp_form_mp}
		\min_{\bm{y}} \; &  Z(\bm{y}) \\
		\text{s.t.} \;& \bm{K}\bm{y} - \bm{b} \leq 0, \\
		& \bm{y} \in \{0,1\}^m.
	\end{aligned}
\end{equation}

Since $Z(\bm{y})$ is not known explicitly, it is approximated via Benders optimality cuts. The $k$th optimality cut, denoted by $\mathcal{C}^{(k)}$, is given by
\begin{equation}\label{eq:optimality_cut}
	\Theta_b \geq f(\bm{x}^k) + \bm{e}^\top\hspace{-1mm} \bm{y}
	+ \bm{\lambda}_k^\top\hspace{-1mm}\left( \bm{h}(\bm{x}^k) + \bm{A}\bm{y}\right)
	+ \bm{\mu}_k^\top \hspace{-1mm}\left( \bm{g}(\bm{x}^k) + \bm{B}\bm{y} \right)\hspace{-1mm},
\end{equation}
where \( \bm{x}^k \) is an optimal solution of $\mathcal{S}(\bm{y}^k)$, and \( \bm{\lambda}_k \), \( \bm{\mu}_k \) are the corresponding dual multipliers. We denote the right-hand side (RHS) of $\mathcal{C}^{(k)}$ by
\[
\mathcal{O}_k(\bm{y}) \coloneqq f(\bm{x}^k) + \bm{e}^\top\hspace{-1mm} \bm{y}
+ \bm{\lambda}_k^\top\hspace{-1mm}\left( \bm{h}(\bm{x}^k) + \bm{A}\bm{y}\right)
+ \bm{\mu}_k^\top\hspace{-1mm} \left( \bm{g}(\bm{x}^k) + \bm{B}\bm{y} \right)\hspace{-1mm}.
\]
The master problem $\mathcal{M}$ is defined as
\begin{equation}
	\begin{aligned}
		\min_{\Theta_b,\, \bm{y}} \; & \Theta_b \\
		\text{s.t.} \quad 
		& \mathcal{C}^{(k)}, \quad \forall k \in K_O, \\
		& \bm{K}\bm{y} - \bm{b} \leq 0, \\
		& \bm{y} \in \{0,1\}^m
	\end{aligned}
	\label{eq:master}
\end{equation}
where \(K_O\) denotes the set of optimality cuts accumulated up to iteration $k$. The solution of $\mathcal{M}$ provides a lower bound on the optimal value.

\section{Graph-based agent for the master problem}
\begin{figure*}[t]
	\centering
	\includegraphics[scale=0.9]{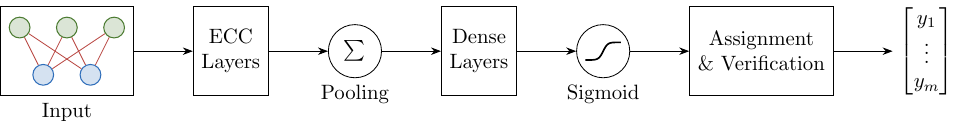}
	\caption{Workflow of the graph-based agent for the master problem.}
	\label{fig:fig_graph_rl}
\end{figure*}

We employ a graph-based RL agent together with a verification mechanism to solve $\mathcal{M}$~\citep{agyeman2026graph}. The overall workflow is shown in Fig.~\ref{fig:fig_graph_rl}. The agent receives a bipartite graph representation of $\mathcal{M}$ and outputs a probability vector over the binary decision variables, which is used to generate candidate binary assignments.

\subsection{Bipartite graph of the master problem}
Each instance of $\mathcal{M}$ is encoded as a bipartite graph \(
\mathcal{G} \coloneqq (V, E, A_d, X_f, X_e),
\) where $V$ is the set of nodes, $E$ is the set of edges, $A_d$ is the adjacency matrix, $X_f$ is the node feature matrix, and $X_e$ is the edge feature matrix. The main components of $\mathcal{G}$ are explained below:

\begin{enumerate}
	\item \textbf{Nodes}: $V$ is partitioned into variable nodes (binary decision variables) and constraint nodes (pure-binary constraints and optimality cuts).
	
	\item \textbf{Node features}: Variable node features correspond to the values of the binary variables from the previous iteration (or an initial assignment at the first iteration), while constraint node features correspond to the RHS values of the associated constraints.
	
	\item \textbf{Edges}: An edge connects a variable node and a constraint node whenever the variable appears in that constraint, and this connectivity is encoded in $A_d$.
	
	\item \textbf{Edge features}: The entries of $X_e$ correspond to the nonzero coefficients linking binary variables and constraints.
\end{enumerate}

\subsection{Agent design and training}
The agent is trained in an actor--critic RL framework, with both actor and critic parameterized by graph neural networks (GNNs). The actor (policy) network takes $(A_d, X_f, X_e)$ as input, applies edge-conditioned convolution (ECC)~\citep{simonovsky2017dynamic}, performs global sum-pooling, and outputs a vector of sigmoid units, one per binary variable. The critic shares the same architecture up to the global sum-pooling layer but uses a single output unit with no activation function to approximate the state-value function.
The Markov decision process elements of the agent are defined as follows:

\begin{enumerate}
	\item \textbf{State}: The graph $\mathcal{G}$ of the current instance of $\mathcal{M}$.

	\item \textbf{Action}\label{sec:action}: The action $\bm{a} \in \{0,1\}^m$ is obtained by sampling independently from Bernoulli distributions parameterized by the policy output:
	\begin{equation} \label{eq:action_selection}
		\begin{aligned}
			\pi_{\theta}(A_d, X_f, X_e) &\coloneqq \bm{p} = (p_1,\dots,p_m), \\
			a_i &\sim \text{Bernoulli}(p_i), \quad i = 1,\dots,m, \\
			\bm{a} &\coloneqq (a_1,\dots,a_m).
		\end{aligned}
	\end{equation}

	\item \textbf{Transition dynamics}: At iteration $k$, the policy uses $(A_{d_k}, X_{f_k}, X_{e_k})$ to produce $\bm{a}_k$ via Eq.~\eqref{eq:action_selection}. If $\bm{a}_k$ satisfies the pure-binary constraints, it is passed to the subproblem; otherwise, $\mathcal{M}$ is solved using an MIP solver to obtain a feasible assignment $\overline{\bm{y}}_k$. Applying $\bm{a}_k$ or $\overline{\bm{y}}_k$ to the subproblem yields an optimality cut, whose addition updates $\mathcal{M}$ and its graph representation from $(A_{d_k}, X_{f_k}, X_{e_k})$ to $(A_{d_{k+1}}, X_{f_{k+1}}, X_{e_{k+1}})$.
	
	\item \textbf{Reward design}: The agent is rewarded for producing feasible assignments, improving the bound gap, and reducing the subproblem solution time $t_{\text{SP}}$. The scalar reward is
	\begin{equation}
		r \coloneqq \alpha_1 r_{\text{feas}} + \alpha_2 r_{\text{gap}} - \alpha_3 r_{\text{time}},
	\end{equation}
	with
	\begin{equation}
		r_{\text{feas}} \coloneqq
		\begin{cases}
			-\beta_1, & \text{if infeasible}, \\[2pt]
			\beta_2,  & \text{if feasible},
		\end{cases}
	\end{equation}
	\begin{equation}\label{eq:bound_improvement}
		r_{\text{gap}} \coloneqq
		\begin{cases}
			\displaystyle \max\!\left(0,\ \frac{\Delta_{k-1}-\Delta_k}{\Delta_0}\right) & \text{if $\mathcal{M}$ is feasible}, \\
			0 & \text{otherwise},
		\end{cases}
	\end{equation}
	\begin{equation}
		r_{\text{time}} \coloneqq \min(t_{\text{SP}}, \tau),
	\end{equation}
	where $\Delta_k \coloneqq \mathrm{UBD}_k - \Theta_b^{(k)}$, $\Theta_b^{(k)}$ denotes the candidate lower-bound value at iteration $k$, and $\alpha_1$, $\alpha_2$, $\alpha_3$, $\beta_1$, $\beta_2$, $\tau$ are nonnegative tuning parameters.
\end{enumerate}

\subsection{Verification mechanism}
The graph-based agent is not guaranteed to produce feasible or optimal assignments. In particular, its predictions may violate the constraints of the master problem and, even when feasible, may induce candidate lower-bound values that are not valid lower bounds for the original MINLP. To address these issues, the integration of the agent into GBD is coordinated through a confidence-based post-processing step. A high-confidence region $[0,\delta_1] \cup [\delta_2,1]$ is defined for the policy output $\bm{p} = (p_1,\dots,p_m)$, where variables with $p_i \le \delta_1$ or $p_i \ge \delta_2$ are fixed to $0$ or $1$, while the remaining variables are left free for a mixed-integer programming (MIP) solver. This procedure yields three regimes:

\begin{enumerate}
	\item \textbf{Full assignment}: All variables are fixed, yielding a candidate $\hat{\by}$. The candidate is first checked for feasibility with respect to the pure-binary constraints. If any constraint is violated, $\hat{\by}$ is discarded and $\mathcal{M}$ is solved using an MIP solver. Otherwise, the candidate lower-bound value associated with $\hat{\by}$ is computed as
	\[
	\hat{\Theta}_b \coloneqq \max_{k \in K_O} \mathcal{O}_k(\hat{\by}).
	\]
	The candidate value $\hat{\Theta}_b$ is accepted only if $\hat{\Theta}_b \le \mathrm{UBD}$; otherwise, $\hat{\by}$ is discarded and $\mathcal{M}$ is solved using an MIP solver.
	
\item \textbf{Partial assignment}: A subset of variables is fixed and $\mathcal{M}$ is solved using an MIP solver with these values fixed. If infeasible, the fixing is discarded and $\mathcal{M}$ is solved with all variables free. If feasible, the resulting solution $(\overline{\by}, \overline{\Theta}_b)$ is obtained, and the corresponding value $\overline{\Theta}_b$ is treated as a candidate lower-bound value. The solution $(\overline{\by}, \overline{\Theta}_b)$ is accepted if $\overline{\Theta}_b \le \mathrm{UBD}$; otherwise, $\mathcal{M}$ is re-solved with all variables free.
	
	\item \textbf{No assignment}: No variable is fixed and $\mathcal{M}$ is solved directly with all variables free.
\end{enumerate}

Since the values $\hat{\Theta}_b$ and $\overline{\Theta}_b$ are not guaranteed to be valid lower bounds of the original MINLP, they are treated as candidate values. A record of these values is maintained to define a surrogate termination criterion based on the gap between the incumbent UBD and the candidate values.

Unlike classical GBD, where the LBD sequence is nondecreasing, the agent-based approach does not guarantee this property. To enforce a monotone sequence, we define the candidate lower bound (CLBD) as
\[
\mathrm{CLBD}_k \coloneqq \max\{\Theta_b^{(k)}, \mathrm{CLBD}_{k-1}\},
\]
where $\Theta_b^{(k)} \in \{\hat{\Theta}_b, \overline{\Theta}_b\}$ is the candidate value at iteration $k$. The algorithm is terminated when the gap between the incumbent UBD and $\mathrm{CLBD}_k$ falls below a specified tolerance.

\section{KKT-informed subproblem predictor}
\subsection{Learning task design}
In classical GBD, the Benders optimality cut \eqref{eq:optimality_cut} is constructed by solving $\mathcal{S}(\bm{y}^k)$ for a given $\bm{y}^k$ at each iteration $k$. This problem can be characterized via the KKT conditions:
\begin{equation}
	\begin{aligned}
		\nabla f(\bm{x}^k) + \bm{\mu}_k^\top \nabla \bm{g}(\bm{x}^k) + \bm{\lambda}_k^\top \nabla \bm{h}(\bm{x}^k) = 0, \\
		\bm{h}(\bm{x}^k) + \bm{A} \bm{y}^k = 0, \\
		\bm{g}(\bm{x}^k) + \bm{B} \bm{y}^k \leq 0, \\
		\bm{\mu}_k^\top \bigl(\bm{g}(\bm{x}^k) + \bm{B} \bm{y}^k\bigr) = 0, \\
		\bm{\mu}_k \geq 0.
	\end{aligned}
\end{equation}

To reduce the computational burden associated with repeatedly solving the subproblem, we introduce a KKT-informed neural network (KINN) predictor. Given $\bm{y}^k$, the model $\text{KINN}_{\omega}(\cdot)$ predicts approximate primal and dual variables
\[
\text{KINN}_\omega(\bm{y}^k) \coloneqq (\hat{\bm{x}}^k, \hat{\bm{\mu}}_k, \hat{\bm{\lambda}}_k),
\]
where $\omega$ denotes the network parameters. The model is trained using a KKT-informed loss that minimizes residuals of the stationarity ($\ell_{\text{stat}}$), primal feasibility ($\ell_{\text{pri}}$), and complementarity ($\ell_{\text{comp}}$) conditions:
\begin{align}
	& \ell_{\text{stat}} = \big\|\nabla f(\hat{\bm{x}}^k) + \hat{\bm{\mu}}_k^\top \nabla \bm{g}(\hat{\bm{x}}^k) + \hat{\bm{\lambda}}_k^\top \nabla \bm{h}(\hat{\bm{x}}^k) \big\|_2^2, \\
	& \ell_{\text{pri}} = \|\max(0, \bm{g}(\hat{\bm{x}}^k)\hspace{-0.5mm} +\bm{B}\bm{y}^k)\|_2^2
	+ \|\bm{h}(\hat{\bm{x}}^k)+\bm{A}\bm{y}^k\|_2^2, \\
	& \ell_{\text{comp}} = \sum_{i=1}^q \phi_\varepsilon\!\left(\hat{\mu}_i,\,-g_i(\hat{\bm{x}}^k)-\bm{B}_i \bm{y}^k\right)^2,
\end{align}
where $\phi_\varepsilon(a,b)=a+b-\sqrt{a^2+b^2+\varepsilon^2}$ is the smoothed Fischer--Burmeister function with smoothing parameter $\varepsilon$. Dual feasibility is enforced via the softplus function $\hat{\bm{\mu}}_k \leftarrow \ln(1+e^{\hat{\bm{\mu}}_k})$. The overall loss is
\[
\ell = \alpha_{\text{sp}} \ell_{\text{stat}} + \beta_{\text{sp}} \ell_{\text{pri}} + \gamma_{\text{sp}} \ell_{\text{comp}},
\]
with positive weights $\alpha_{\text{sp}}, \beta_{\text{sp}}, \gamma_{\text{sp}}$.

\begin{figure}[t]
	\centering
	\includegraphics[
	width=0.8\columnwidth,
	trim=1.5cm 0 0 0,
	]{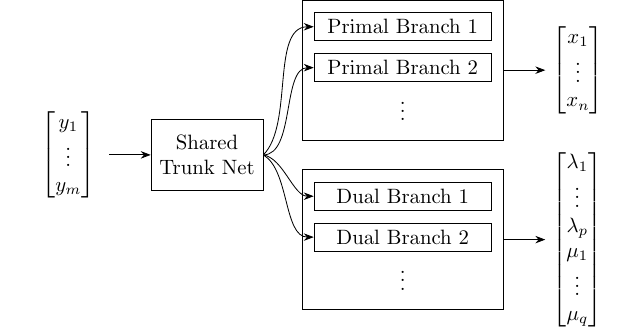}
	\caption{The architecture of the proposed KINN.}
	\label{fig:pinn_diag}
\end{figure}

To improve prediction performance, we employ a branched architecture that separately models the primal and dual variables, as illustrated in Fig.~\ref{fig:pinn_diag}. A shared trunk extracts common features from $\bm{y}^i$, followed by branch networks that predict the primal $\hat{\bm{x}}^i$ and dual $(\hat{\bm{\mu}}_i, \hat{\bm{\lambda}}_i)$ variables.

For training, we construct a dataset $\mathcal{Y} = \{\bm{y}^i\}_{i=1}^N$ of feasible binary assignments satisfying the pure-binary constraints. The model is trained using mini-batch stochastic gradient descent, as presented in Algorithm~\ref{alg:nn-training}.

\begin{algorithm}[t]
	\caption{KINN training procedure}
	\label{alg:nn-training}
	\begin{algorithmic}[1]
		\Require $\mathcal{Y} = \{\by^i\}_{i=1}^{N}$,
		$\text{KINN}_\omega$,
		$\ell$,
		epochs $T$, batch size $B$, learning rate $\eta$
		\For{$t = 1, 2, \dots, T$}
		\For{each mini-batch $\mathcal{Y}_B \subset \mathcal{Y}$ of size $B$}
		\State For each $\by^i \in \mathcal{Y}_B$, compute $(\hat{\bx}^i, \hat{\bm{\mu}}_i, \hat{\blam}_i) \gets \text{KINN}_\omega(\by^i)$
		\State Evaluate
		\(
		L(\omega) \gets \frac{1}{|\mathcal{Y}_B|}
		\sum_{\by^i \in \mathcal{Y}_B} \ell(\by^i, \hat{\bx}^i, \hat{\bm{\mu}}_i, \hat{\blam}_i)
		\)
		\State $g \gets \nabla_\omega L(\omega)$
		\State $\omega \gets \omega - \eta \, g$
		\EndFor
		\EndFor
		\State \Return $\omega$
	\end{algorithmic}
\end{algorithm}

\subsection{Deployment}
During deployment, the trained KINN is used to predict approximate primal--dual solutions for a given $\bm{y}^k$. These predictions are used to construct Benders optimality cuts without explicitly solving the subproblem. As the predicted solutions do not guarantee exact satisfaction of the KKT conditions, the resulting cuts are inexact. The impact of these inexact cuts on GBD convergence is assessed empirically.

\section{Case study}\label{sec:simulations}
We consider an MINLP adapted from~\citep{floudas1995nonlinear}, with parameterized objective coefficients:
\begin{align}
	\min_{\bm{x},\bm{y}} \quad &
	c_1y_{1} + c_2y_{2} + c_3y_{3} + c_4y_{4} + c_5y_{5} 
	- 10x_{3} - 15x_{5} - 15x_{9} \nonumber\\[-2.5mm]
	& + 15x_{11} + 5x_{13} - 20x_{16} 
	+ \exp(x_{3}) + \exp\!\left(\frac{x_{5}}{1.2}\right)\nonumber\\
	& - 60\ln(x_{11} + x_{13} + 1) + 140 \tag{E1} \label{E1}
\end{align}
\vspace{-4mm}
\begin{align}
	\text{s.t.} \quad
	& -\ln(x_{11} + x_{13} + 1) \le 0 \tag{E2} \label{E2} \\
	& - x_{3} - x_{5} - 2x_{9} + x_{11} + 2x_{16} \le 0 \tag{E3} \label{E3} \\
	& - x_{3} - x_{5} - 0.75x_{9} + x_{11} + 2x_{16} \le 0 \tag{E4} \label{E4} \\
	& x_{9} - x_{16} \le 0 \tag{E5} \label{E5} \\
	& 2x_{9} - x_{11} - 2x_{16} \le 0 \tag{E6} \label{E6} \\
	& -0.5x_{11} + x_{13} \le 0 \tag{E7} \label{E7} \\
	& 0.2x_{11} - x_{13} \le 0 \tag{E8} \label{E8} \\
	& \exp(x_{3}) - U y_{1} \le 1 \tag{E9} \label{E9} \\
	& \exp\!\left(\frac{x_{5}}{1.2}\right) - U y_{2} \le 1 \tag{E10} \label{E10} \\
	& 1.25x_{9} - U y_{3} \le 0 \tag{E11} \label{E11} \\
	& x_{11} + x_{13} - U y_{4} \le 0 \tag{E12} \label{E12} \\
	& -2x_{9} + 2x_{16} - U y_{5} \le 0 \tag{E13} \label{E13} \\
	& y_{1} + y_{2} = 1 \tag{E14} \label{E14} \\
	& y_{4} + y_{5} \le 1 \tag{E15} \label{E15}
\end{align}
\vspace{-4mm}
\begin{align*}
	& \bm{y}^{\top} = (y_{1},y_{2},y_{3},y_{4},y_{5}) \in \{0,1\}^{5}, \\
	& \bm{x}^{\top} = (x_{3},x_{5},x_{9},x_{11},x_{13},x_{16}) \in \mathbb{R}^{6}, \\
	& \bm{a}^{\top} = (0,0,0,0,0,0), \quad 
	\bm{b}^{\top} = (2,2,2,\text{--},\text{--},3), \\
	& \bm{a} \le \bm{x} \le \bm{b}, \quad U = 10.
\end{align*}

The coefficients $c_1,\dots,c_4$ are sampled as positive integers from $[1,39]$, while $c_5$ is sampled from $[1,7]$. This parameterization enables the generation of diverse problem instances for training and evaluation.

\subsection{Graph-based agent development}
The actor network comprised two ECC layers with 64 units and ReLU activations, followed by global sum pooling, a dense layer with 64 ReLU units, and an output layer with 5 sigmoid units. The critic shared the same architecture up to the global pooling layer but employed a single output unit with no activation function to approximate the value function. Prior to RL training, behavioural cloning was used to obtain a warm-start policy for the actor. Specifically, 3{,}000 instances were solved using GBD, with Gurobi~12.0.1~\citep{gurobi} for the master problem and IPOPT~3.12.6~\citep{wachter2006ipopt} for the subproblem, and the actor was trained to imitate the resulting optimal binary assignments.

The agent was trained over 10{,}000 episodes using proximal policy optimization (PPO)~\citep{schulman2017proximal}. The training hyperparameters were: batch size $=32$, epochs $=5$, learning rate $=5\times10^{-4}$, discount factor $=0.99$, generalized advantage estimation parameter $=0.97$, and clipping parameter $=0.25$. The reward parameters were set to $\alpha_1 = 1.0$, $\alpha_2 = 6.0$, $\alpha_3 = 1.0$, $\tau = 1.0$, $\beta_1 = 0.50$, and $\beta_2 = 1.50$. The termination criterion was $1\times10^{-3}$, and the confidence thresholds were $\delta_1 = 0.10$, $\delta_2 = 0.90$.

\subsection{KINN development}
The KINN architecture consisted of a shared trunk network and three branch networks. A single primal branch (PB) predicted the primal variables $\bx$. For the dual variables, one branch (DB1) predicted the dual variables associated with constraints \eqref{E9}--\eqref{E13} involving both $\bx$ and $\by$, while a second branch (DB2) predicted the remaining dual variables associated with constraints involving only $\bx$. The trunk network consisted of two layers with 64 neurons each. Each branch network consisted of two layers with 64 and 128 neurons, respectively. The input dimension was 5, and the output dimensions of PB, DB1, and DB2 were 6, 5, and 17, respectively.

In the training phase, the set of complicating variables $\mathcal{Y} = \{\by^i\}_{i=1}^{12}$ was generated from the pure binary constraints \eqref{E14} and \eqref{E15}. Due to the small size of $\mathcal{Y}$, full-batch training was used at each epoch. The weights in the KKT-informed loss were set to $\alpha_\text{sp}=\beta_\text{sp}=\gamma_\text{sp}=1$ and the smoothing parameter $\varepsilon$ in $\phi_{\varepsilon}(\cdot)$ was set to $1\times10^{-3}$. The model was trained for 40{,}000 epochs, followed by fine-tuning of the dual branches for an additional 40{,}000 epochs with the trunk and primal branch frozen. The learning rate was set to $1\times10^{-4}$.

\subsection{Performance evaluation}
A comparative study over 100 test instances was conducted to evaluate the proposed approach against three baselines. The first baseline corresponded to classical GBD, in which the master problem was solved using BONMIN 1.8.9~\citep{bonami2008bonmin} and the subproblem using IPOPT. The second baseline (agent-only) replaced the master problem solver with the trained graph-based agent, while the subproblem was solved using IPOPT. The third baseline (KINN-only) replaced the subproblem solver with the trained KINN, while the master problem was solved using BONMIN. The proposed approach combined both surrogates.

The methods were compared using the average total solution time per instance ($\bar{t}_{\text{tot}}$), average master problem time ($\bar{t}_{\text{MP}}$), average subproblem time ($\bar{t}_{\text{SP}}$), and average number of iterations ($\bar{N}_{\text{iter}}$). The reference solution for each instance was obtained by directly solving the original MINLP using BONMIN. All experiments were conducted on a desktop equipped with an Intel i7-14700 CPU and 32\,GB RAM.

Table~\ref{tab:perf_summary} shows that the single-surrogate approaches achieved reductions in total solution time relative to classical GBD, with decreases of 50.8\% (agent-only) and 14.0\% (KINN-only). The proposed hybrid approach achieved the largest reduction of 57.5\%. All approaches converged in a comparable number of iterations, indicating that the observed speedups arose primarily from reduced per-iteration computational cost. All methods recovered the reference solution across the 100 test instances.

Table~\ref{tab:kkt_quality} reports the predictive performance of the KINN. The NRMSE values indicate accurate prediction of both the primal and dual variables. The stationarity and primal feasibility residuals are on the order of $10^{-3}$, showing that the predicted solutions approximately satisfy the KKT conditions. The complementarity residual is comparatively larger, reflecting the difficulty of enforcing complementarity conditions during training. Despite these approximation errors, the empirical results indicate that the resulting inexact cuts do not degrade the convergence behavior of GBD, as  the reference solution was recovered in all the test instances. These results suggest that the KINN provides sufficiently accurate approximations for practical use within the GBD framework.
\begin{table*}[t]
	\centering
	\caption{Summary of performance over 100 test instances. CPU times are reported as mean $\pm$ std per instance; percentage reduction relative to classical GBD is shown in parentheses.}
	\label{tab:perf_summary}
	\begin{tabular}{lcccc}
		\toprule
		Approach
		& $\bar{t}_{\text{tot}}$ [s] & $\bar{t}_{\text{MP}}$ [s] & $\bar{t}_{\text{SP}}$ [ms] & $\bar{N}_{\text{iter}}$ [--] \\
		\midrule
		Classical GBD & $0.64\pm0.14$          & $0.56\pm0.14$ & $71.2\pm7.3$ & $9.2\pm0.9$ \\
		Agent-only    & $0.32\pm0.11$ (50.8\%) & $0.23\pm0.10$ & $71.9\pm7.2$ & $9.3\pm0.9$ \\
		KINN-only     & $0.55\pm0.12$ (14.0\%) & $0.54\pm0.12$ & $0.5\pm0.1$  & $9.1\pm0.9$ \\
		Proposed      & $0.27\pm0.08$ (57.5\%) & $0.26\pm0.08$ & $0.4\pm0.1$  & $9.2\pm0.9$ \\
		\bottomrule
	\end{tabular}
\end{table*}
\begin{table*}[ht]
	\centering
	\caption{KINN accuracy over 100 test instances ( residuals are reported as mean $\pm$ std).}
	\label{tab:kkt_quality}
	\begin{tabular}{ccccc}
		\toprule
		NRMSE (primal) & NRMSE (dual)
		& Stationarity residual & Primal feas.\ residual & Complementarity residual \\
		\midrule
		$1.20\times10^{-3}$ & $6.60\times10^{-2}$
		& $6.78\times10^{-3}\pm7.76\times10^{-3}$
		& $3.52\times10^{-3}\pm4.02\times10^{-3}$
		& $1.50\times10^{-1}\pm1.72\times10^{-1}$ \\
		\bottomrule
	\end{tabular}
\end{table*}
\section{Conclusion}
This paper presents a hybrid learning-augmented framework for accelerating GBD, in which a graph-based agent guides the solution of the master problem and a KKT-informed neural network approximates the solution of the subproblem. A confidence-based assignment mechanism is employed to ensure a robust integration of the agent’s policy into the GBD algorithm. On a representative MINLP case study, the proposed approach achieves a substantial reduction in total solution time relative to classical GBD. The results further indicate that the use of inexact cuts generated by the KINN does not degrade convergence in the tested instances. These findings demonstrate the effectiveness of integrating learning-based surrogates within decomposition algorithms. Future work will investigate solver-assisted strategies to further improve subproblem approximations and enhance the robustness of the proposed framework.
\bibliography{references}
\end{document}